\documentclass[aps,prl,showpacs,superscriptaddress,amssymb,twocolumn]{revtex4-1}
\usepackage{graphicx}
\usepackage{appendix} 
\usepackage{bm,amsmath}
\usepackage{dcolumn}

\begin{document}

\rightline{\tt Phys.~Rev.~Lett.~{\bf 127},~090502~(2021)}

\vspace{0.2in}

\title{Conditionally rigorous mitigation of multiqubit measurement errors}

\author{Michael R. Geller}
\affiliation{Center for Simulational Physics, University of Georgia, Athens, Georgia 30602, USA}

\date{September 9, 2021}

\begin{abstract}
Several techniques have been recently introduced to mitigate errors in near-term quantum computers without the overhead required by quantum error correcting codes. While most of the focus has been on gate errors, measurement errors are significantly larger than gate errors on some platforms. A widely used {\it transition matrix error mitigation} (TMEM)
technique uses measured transition probabilities between initial and final classical states to correct subsequently measured data. However from a rigorous perspective, the noisy measurement should be calibrated with perfectly prepared initial states and the presence of any state-preparation error corrupts the resulting mitigation. Here we develop a measurement error mitigation technique, conditionally rigorous TMEM,  that is not sensitive to state-preparation errors and thus avoids this limitation. We demonstrate the importance of the technique for high-precision measurement and for quantum foundations experiments by measuring Mermin polynomials on IBM Q superconducting qubits. An extension of the technique allows one to correct for both state-preparation and measurement (SPAM) errors in expectation values as well; we illustrate this by giving a protocol for fully SPAM-corrected quantum process tomography.
\end{abstract}

\maketitle

There is a large effort focused on the development of practical applications for near-term quantum computers \cite{HarrowNat17,180100862,190401502,AruteNat19}, for which effective error characterization \cite{190208543,190712976,190713022,200601805} and mitigation \cite{161109301,TemmePRL17,EndoPRX18,DumitrescuPRL17,180504492,SongSciAdv19,TannuIEEE19,TannuIEEE19b,OttenPRA19,200312314,200510189,200703663,201007496,201009188,DewesPRL12,160304512,180411326,190411935,190708518,191001969,191113289,181010523,200614044,201008520,BialczakNatPhys10,NeeleyNat10,181102292} are essential.  Also interesting is the use of near-term quantum computers for studying foundational problems in quantum mechanics and quantum information science, such verifying entanglement \cite{BialczakNatPhys10,NeeleyNat10,181102292,DiCarloNat10,AlsinaPRA16,171205642,200511271,200512504,MonzPRL11,LanyonPRL14,SongSci19,190505720}, finding consistent histories \cite{ArrasmithNatComm19}, and observing information scrambling \cite{LiPRX17,180602807,200303307}, for which high-precision measurement is also required.

In gate-based quantum computers, where errors are separated into state-preparation errors, gate errors, and measurement errors, measurement errors are often the largest, and they can increase with register size due to persistent crosstalk \cite{181010523,200601805,200614044}. A well-known technique for mitigating these errors is to measure the matrix $T$ of transition probabilities between all initially prepared and observed classical states $x \in \{0,1\}^n$ on an empty (identity) circuit, and then minimize $\| T \, p_{\rm corr} - p_{\rm noisy} \|_2^2$ subject to constraints $0 \le p_{\rm corr}(x) \le 1$ and $\| p_{\rm corr} \|_1 =1$ to correct subsequently measured probability distributions \cite{DewesPRL12,160304512,180411326,190411935,190708518,191001969,191113289,181010523,200614044,201008520,BialczakNatPhys10,NeeleyNat10,181102292}. Here $T$ is a $2^n \times 2^n$ stochastic matrix with elements
\begin{equation}
T(x|x') = {\rm Pr}(x|x') ={\rm tr} (E_x \, \rho_{x'}),
\label{defT}
\end{equation}
where $\{ E_x\}_{x \in \{0,1\}^n }$ is the noisy multiqubit POVM,
$\rho_{x'}$ is an initially prepared state, ideally equal to the classical state $|x'\rangle \langle x' |$, and $n$ is the number of qubits in the processor or active register. Each column of $T$ is the raw conditional probability distribution ${\rm Pr}(x|x')$ measured immediately after preparing $x' \! .$ $p_{\rm noisy}$ is a given measured probability distribution expressed as a vector, and $ p_{\rm corr}$ is the corrected distribution. $\| \cdot  \|_2$ is the Euclidean norm and $\| \cdot  \|_1$ is the $\ell_1$-norm. An implementation of this {\it transition matrix error mitigation} (TMEM)
technique is available in qiskit \cite{QiskitShort}, IBM Q's software development kit, and TMEM variations based on maximum likelihood estimation \cite{190411935} and iterative Bayesian unfolding \cite{191001969} have also been investigated. 

Despite the wide use of TMEM, its theoretical justification has only recently been investigated \cite{190708518,200201471}. This work showed that imperfect measurements described by POVMs that are strictly diagonal in the classical basis---representing a type of biased classical noise---can be exactly corrected (up to statistical errors) by following the protocol described above but with a different matrix, which we write as
\begin{equation}
\Gamma(x|x') := {\rm tr} (E_x \, |x'\rangle \langle x' |) = \langle x' | E_x | x' \rangle,
\label{defGamma}
\end{equation}
which has no state-preparation error. Reconstructing noisy measurement via detector tomography \cite{ LuisPRL99}, which also assumes perfect state preparation, leads to the same conclusion. But how can we obtain $\Gamma$ in the presence of inevitable state-preparation errors?

The need to prepare accurate classical states is a frequent requirement in 
 experimental quantum computation, beyond the example (\ref{defGamma}). An extension of our technique can be used to correct other expectation values for both state-preparation and measurement (SPAM) errors. An example of independent interest, SPAM-corrected quantum process tomography, is given at the end of the paper.
 
In \cite{200201471} we calculated $\Gamma$ for individual superconducting qubits by using single-qubit gate-set tomography (GST) \cite{BlumeKohout13104492,MerkelPRA13,BlumeKohoutNat17}, which simultaneously estimated a prepared state $\rho_0 \! \approx  \!  |0\rangle \langle 0 |$, the $\pi/2$ rotations $G_{\rm x}  \!  \approx  \!  e^{-i (\pi/4) X}$ and $G_{\rm y}  \!  \approx  \!  e^{-i (\pi/4) Y}$, and the 2-outcome POVM elements $E_0  \! \approx  \!  |0\rangle \langle 0 |$ and $E_1 = I - E_0.$ Here $X$ and $Y$ are Pauli matrices and $I$ is the identity. The $\approx$ symbol means that, when the errors are small, the noisy quantities are close to the indicated targets. The estimated $E_{0,1}$ were found to be nearly diagonal, consistent with the expectation that the dominant source of measurement error in transmon qubits is $T_1$ relaxation during dispersive readout \cite{MalletNatPhys09,180107904}. The resulting $\Gamma$ matrices, obtained from the estimated POVM, were found to be significantly different than the concurrently measured $T$ matrices. But extending this approach beyond one or two qubits is not practical due to the high sample complexity of multiqubit GST. 

In this work we introduce and demonstrate a technique to estimate $\Gamma$ by combining TMEM with {\it single-qubit} GST for each qubit in the register. Our approach assumes that the prepared classical states are separable, i.e., we neglect {\it entangling} crosstalk errors during state preparation (this restriction is lifted below 
after making additional locality assumptions). In this case knowledge of the noisy $\rho_{x'}$ on each qubit, as estimated by GST, also specifies a particular linear combination of noisy initial states that is equivalent to each {\it ideal} classical state $ |x'\rangle \langle x'|$, up to statistical errors. Our technique combines ideas  from TMEM \cite{DewesPRL12,160304512,180411326,190411935,190708518,191001969,191113289,181010523,200614044,201008520,BialczakNatPhys10,NeeleyNat10,181102292}, tomography \cite{BlumeKohout13104492,MerkelPRA13,BlumeKohoutNat17}, and quasiprobability decompositions \cite{TemmePRL17,EndoPRX18,SongSciAdv19}, and can be used whenever a probability distribution is  estimated. 

We now show that (\ref{defGamma}) can be experimentally measured by expressing it as a linear combination of expectation values that are measured with no ideal state-preparation assumption. This is possible because GST reveals the density matrices of the noisy states (a brief introduction to GST and a discussion of its assumptions and limitations are provided in \cite{SI}). First note that any single-qubit state $\rho$ can be written as a unique linear combination of the four ideal projectors
\begin{equation}
\lbrace \pi_{0}, \pi_{1} , \pi_{+},  \pi_{+i} \rbrace
\label{defLambdaBasis}
\end{equation} 
where
\begin{eqnarray}
\pi_{0} &=& \begin{pmatrix} 1 & 0 \\ 0 & 0 \\ \end{pmatrix}, \ \ 
\pi_{1} = \begin{pmatrix} 0 & 0 \\ 0 & 1 \\ \end{pmatrix} , \\
\pi_{+} &=& \frac{1}{2}  \begin{pmatrix} 1& 1 \\ 1 & 1 \\ \end{pmatrix}, \ \ 
\pi_{+i} = \frac{1}{2} \begin{pmatrix} 1 & -i \\ i & 1 \\ \end{pmatrix}.
\end{eqnarray}
The expansion coefficients (quasiprobabilities) are purely real. In terms of $\rho$'s  Bloch vector $\vec r = (x,y,z)$ we have
\begin{equation}
\rho = \textstyle{( \frac{1-x-y+z}{2})} \, \pi_{0} + \textstyle{( \frac{1-x-y-z}{2})} \, \pi_{1} +  x \, \pi_{+}  + y \, \pi_{+i} .
\label{defRhoExpansion}
\end{equation} 

Next we consider, for each qubit in the register, a set of four noisy initial states
\begin{equation}
\lbrace \rho_{0}, \rho_{1}, \rho_{+}, \rho_{+i},  \rbrace,
\label{nonidealStates}
\end{equation} 
where $\rho_\lambda$ is the noisy state prepared after attempting to prepare $\pi_\lambda$, which we will learn from GST. For consistency with what follows, the states $\rho_{1}$, $\rho_{+},$ and $\rho_{+i}$ are to be prepared specifically by applying $G_{\rm x}^2$, $G_{\rm y}$, and $G_{\rm x}^3$ to $ \rho_{0}$, respectively, where
$G_{\rm x} \approx e^{-i (\pi/4) X}$ and $G_{\rm y} \approx e^{-i (\pi/4) Y}$ are $\pi/2$ rotations. Then we have
\begin{equation}
\rho_\lambda = \! \! \sum_{\lambda' \in \{ 0,1,+,+i \} } L_{\lambda  \lambda'} \ \pi_{\lambda'}, \ \ \ {\lambda \in \{ 0,1,+,+i \} } 
\label{defL}
\end{equation}
with the $4 \times 4$ matrix $L$ inferred from GST by using (\ref{defRhoExpansion}). $\rho_{0}$ is directly estimated from GST, and the remaining states are calculated afterwards by acting on $\rho_{0}$ with the estimated noisy gates  $G_{\rm x}$ and $G_{\rm y}.$ In tomographic notation,
\begin{eqnarray}
|\rho_{1} \rangle \rangle &=&  G_{\rm x}^2 \,  |\rho_0 \rangle \rangle \\
|\rho_{+} \rangle \rangle &=&  G_{\rm y} \,  |\rho_0\rangle \rangle \\
|\rho_{+i} \rangle \rangle &=&  G_{\rm x}^3 \,  |\rho_0\rangle \rangle, 
\end{eqnarray}
where $|\rho_\lambda \rangle \rangle$ and $G_\mu \, (\mu\! = \! {\rm x,y})$ are the states and gates in the Pauli basis
\begin{equation}
|\rho_\lambda \rangle \rangle_{\sigma} =  \frac{ {\rm tr} (\rho_\lambda \sigma)}{2} \ \ {\rm and} \ \ (G_\mu)_{\sigma \sigma'} = \frac{ {\rm tr} \, [\sigma \,{\cal G}_\mu(\sigma') ] }{2},
\label{def PTM}
\end{equation}
where  ${\cal G}_\mu$ is the channel representing the noisy $G_\mu$, and $\sigma, \sigma' \in \lbrace I, X, Y Z \rbrace.$

In the error-free case $L$ is equal to the $4\times 4$ identity. If the state-preparation errors are not too large the noisy $L$ can be inverted, yielding, for a single qubit
\begin{equation}
\pi_{\lambda} = \! \! \sum_{\lambda' \in \{ 0,1,+,+i \} } (L^{-1})_{\lambda  \lambda'} \, \rho_{\lambda'} .
\label{defL_1}
\end{equation}
The result (\ref{defL_1}) provides a representation for the ideal single-qubit states (\ref{defLambdaBasis}) in terms of the actual noisy ones, and can be used in other expectation value measurements; we apply (\ref{defL_1})  to process tomography below.

Using (\ref{defL_1}) in (\ref{defGamma}) leads to
\begin{equation}
\Gamma(x|x') = \sum_{\lambda_1 \cdots \lambda_n} 
(L_1^{-1})_{x_1'  \lambda_1} \cdots (L_n^{-1})_{x_n'  \lambda_n}   
\  {\rm tr} (E_x \, \rho_{\lambda_1} \! \cdots \rho_{\lambda_n}   ) ,
\label{GammaEst}
\end{equation}
where $x,x' \in \{0,1\}^n$ are classical states, $x_i'$ is the $i$th bit of $x'$, $L_i$ is the matrix in (\ref{defL}) for qubit $i$, and 
$\rho_{\lambda_1} \! \cdots \rho_{\lambda_n}$ is the $n$-qubit tensor product   
\begin{equation}
\rho_{\lambda_1} \otimes \rho_{\lambda_1} \otimes \cdots \otimes \rho_{\lambda_n}
\end{equation}
of the noisy basis states from (\ref{nonidealStates}). In condensed notation we can write (\ref{GammaEst}) as
\begin{equation}
\Gamma(x|x') \ = \! \!
\sum_{ {\lambda} \in \{ 0,1,+,+i \}^n } \! \!
\big(L^{-1}\big)_{x'  {\lambda} }  
\  {\rm tr} (E_x \, \rho_{{\lambda}} ),
\label{GammaEstCompact}
\end{equation}
where $L = L_1 \otimes L_2 \otimes  \cdots \otimes L_n$.
This expression provides an estimate for $\Gamma$ in terms of the $4^n$ noisy measurements ${\rm tr} (E_x \rho_{{\lambda}}).$ Although the exact $\Gamma$ is necessarily stochastic, sampling errors may result in a slightly non-stochastic estimate ${\hat \Gamma}$, and in these cases we replace ${\hat \Gamma}$ by the stochastic matrix closest in Frobenius distance.

We demonstrate conditionally rigorous TMEM on the IBM Q superconducting processor ibmq\_santiago. This 5-qubit chip with a linear chain geometry was added in 2020 and features a quantum volume of 32, the highest currently available on the IBM Q network. Gate errors, estimated by randomized benchmarking and provided by the backend, are summarized in \cite{SI}. We ran single-qubit GST and calculated the $L$ matrices for qubits $\{ Q_0, Q_1, Q_2, Q_3\}$ using our BQP data acquisition software combined with pyGSTi \cite{pygsti}. We measured 589 distinct circuits up to length 16 (18 including fiducials) on each qubit, and generated robust CPTP estimates for the state $\rho_0,$ the $\pi/2$ rotations $G_{\rm x}$ and $G_{\rm y}$, and the two-outcome POVM. All circuits were measured with 8k measurement samples. The detailed GST results and $L$ matrices are provided in \cite{SI}. Conditionally rigorous TMEM can be independently validated in the $n \! = \! 1$ limit by comparing the single-qubit $\Gamma$ matrix obtained from quasiprobability decompositions, via (\ref{GammaEstCompact}), with that computed directly from the GST-estimated POVM \cite{200201471}. The results of this check confirm the accuracy of (\ref{GammaEstCompact}) in the $n=1$ case and are also provided in \cite{SI}. A classical simulation also validates the technique on a simplified error model consistent with the ibmq\_santiago \cite{SI}.

\begin{figure}
\includegraphics[width=6.0cm]{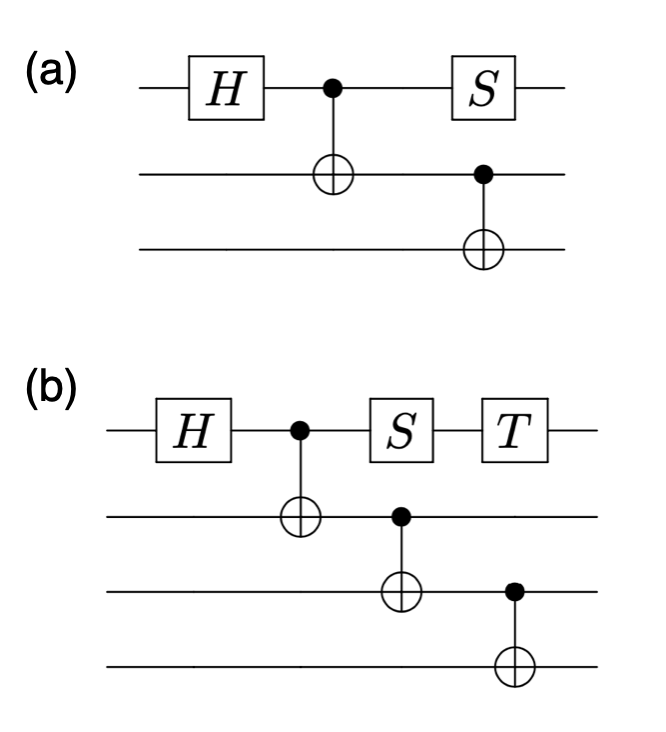} 
\caption{Circuits to prepare (a) the three-qubit state (\ref{defGHZ3}) and (b) the four-qubit state (\ref{defGHZ4}). $H$, $S$, and $T$ are the standard Hadamard, $\pi/2$ phase, and $\pi/4$ phase gates, respectively.}
\label{circuit figure}
\end{figure} 

Next we measure and correct Mermin polynomials \cite{MerminPRL90}. As multipartite generalizations of Bell's inequality, the Mermin inequalities have foundational interest for potentially violating local realism if specific combinations of Pauli expectation values are sufficiently large. Although the presence of detection and locality loopholes prevent us from rejecting local realism without additional assumptions, the accuracy and rigorous justification of any error mitigation technique is clearly important for this experimental test. This example is also interesting because there is a rich history of measuring Mermin polynomials with photons \cite{PanNat00}, superconducting qubits
\cite{NeeleyNat10,DiCarloNat10,AlsinaPRA16,171205642,200511271,200512504} and ions \cite{LanyonPRL14}, enabling comparisons across platforms and over time. Of course any expectation value (or any quantity derived from experimentally estimated probability distributions) can be similarly corrected.
We measure the three-qubit Mermin polynomial
\begin{equation}
M_3 = \langle XXY \rangle + \langle XYX \rangle + \langle YXX \rangle -  \langle YYY \rangle 
\label{defM3}
\end{equation}
on the state 
\begin{equation}
\frac{ |000\rangle + i |111\rangle}{\sqrt 2},
\label{defGHZ3}
\end{equation}
and the four-qubit polynomial 
\begin{eqnarray}
&M_4 = \langle XXXY \rangle + \langle XXYX \rangle + \langle XYXX \rangle +  \langle YXXX \rangle &\nonumber \\
&+ \langle XXYY \rangle + \langle XYXY \rangle + \langle XYYX \rangle +  \langle YXXY \rangle &\nonumber \\
&+ \langle YXYX \rangle + \langle YYXX \rangle - \langle XXXX \rangle -  \langle XYYY \rangle& \nonumber \\
&- \langle YXYY \rangle - \langle YYXY \rangle - \langle YYYX \rangle -  \langle YYYY \rangle \ \ &
\label{defM4}
\end{eqnarray}
on the state
\begin{equation}
\frac{ |0000\rangle + e^{3 \pi i /4}  |1111\rangle}{\sqrt 2}.
\label{defGHZ4}
\end{equation}
The states (\ref{defGHZ3}) and (\ref{defGHZ4}) are chosen because they allow for a maximal violation of local realism in the ideal limit. They are prepared by the circuits shown in Fig.~\ref{circuit figure}. Measurement of $X$ and $Y$ are obtained by applying $H$ and $HS^\dagger$ prior to a $Z$ basis measurement.  We measure every Pauli expectation value in (\ref{defM3}) and (\ref{defM4}) and do not assume that the noisy states have the symmetries under qubit exchange possessed by their ideal targets. 

\begin{table}[htb]
\centering
\caption{Mermin polynomials (\ref{defM3}) and (\ref{defM4}) measured on the IBM Q device ibmq\_santiago with 8k measurement samples. LR is the maximum value consistent with local realism. QM is the maximum value allowed by quantum mechanics. Experimental data are reported as the average over $N\!=\! 16$ independent estimations plus/minus the standard error $\sigma / \sqrt{N}$, where $\sigma^2$ is the variance of the $N$ samples. The $T$ and $\Gamma$ columns give the polynomial values after correcting the raw probability distributions with the $T$ and $\Gamma$ matrices. 
}
\begin{tabular}{|c|c|c|c|c|c|}
\hline
 & LR & QM &  Raw data & $T$ matrix & $\Gamma$ matrix \\
\hline
$M_3$ &2 & 4 & $ 3.618 \pm 0.004 $ & $3.998  \pm 0.002$ & $3.814 \pm 0.004$  \\
\hline 
$M_4$ & 4 & $8\sqrt{2}$ & $8.797 \pm 0.212$ & $10.232  \pm 0.247 $  & $9.037 \pm 0.218$ \\
\hline 
\end{tabular}
\label{mermin table}
\end{table}

The experimental results are summarized in Table \ref{mermin table}. Polynomial $M_3$ was measured on qubits $\{ Q_0, Q_1, Q_2 \}$ after preparing (\ref{defGHZ3}), whereas $M_4$ was measured on qubits $\{ Q_0, Q_1, Q_2, Q_3 \}$ after preparing (\ref{defGHZ4}). We find that in both cases the effects of measurement errors are significant, as reflected in the differences between the raw and rigorously corrected values. We also find that the $T$ matrix {\it overestimates} the entanglement, which means it should not be trusted in quantum foundations experiments such as this one. The $T$ matrix result for $M_3$ is unphysical as it nearly saturates the quantum upper bound, yet also includes two CNOT gate errors of magnitude $0.008$ and $0.007$ \cite{SI}. The differences between correction by $T$ and $\Gamma$ is statistically significant for both polynomials. The complete data set and comparison with previous experiments on superconducting qubits is provided in \cite{SI}.

\begin{table}[htb]
\centering
\caption{Classically simulated measurement and mitigation of the 4-qubit Mermin polynomial. Simulation results are reported as the average over $N\!=\! 16$ random POVM samples. 
}
\begin{tabular}{|c|c|c|c|c|c|c|}
\hline
 & LR & QM & Exact$(\eta \!=\! 0.2)$ & Raw data & $T$ matrix & $\Gamma$ matrix \\
\hline
$M_4$ & 4 & $8\sqrt{2}$ & 9.051 & 8.962 & 9.821& 9.067  \\
\hline 
\end{tabular}
\label{mermin simulation table}
\end{table}

To test the accuracy of rigorous TMEM, we summarize in Table \ref{mermin simulation table} the results of a purely classical simulation of  conditionally rigorous $M_4$ mitigation, assuming a simplified error model designed to be similar to ibmq\_santiago \cite{SI}. Here $\eta$ is the error strength of a depolarization channel applied to (\ref{defGHZ4}) to account for imperfect GHZ state preparation. In the simulation, the multiqubit POVM  consists of diagonal but otherwise random positive semidefinite matrices. Importantly, the measurement operators are {\it not} separable; they contain correlated multiqubit measurement errors. As the measurement-error-free value of $M_4$ is known in this simulation (the ``Exact" column in Table \ref{mermin simulation table}), the accuracy of conditionally rigorous TMEM can be tested. We find that GST successfully learns the simplified error model and that the $\Gamma$ matrix correctly mitigates the correlated measurement errors \cite{SI}.  However we note that rigorous TMEM can fail if GST does (see  \cite{SI} for additional discussion).

In the entanglement estimation above, TMEM is used to remove measurement errors from the Mermin polynomials. We accept the entangled states actually produced in the noisy device, and estimate their entanglement using $\Gamma$-based TMEM, but we do not (and should not) attempt to correct for the imperfect entangled state preparation. However conditionally rigorous TMEM can be extended to the estimation of expectation values with both initial states and measurements corrected: Here we give an application to fully SPAM-corrected quantum process tomography (QPT). Specifically, we consider the estimation of the Pauli transfer matrix (\ref{def PTM}) for an arbitrary CPTP channel $\rho \mapsto \rho' =  \Phi(\rho)$, defined as
\begin{equation}
\Phi_{\sigma \sigma'} := \frac{ {\rm tr} \, [\sigma \, \Phi(\sigma') ] }{d}.
\label{def general PTM}
\end{equation}
Here $\sigma, \sigma' \in \lbrace I, X, Y Z \rbrace^n$ and $d=2^n$ is the Hilbert space dimension. The measurement of observable $\sigma$ is obtained following the standard method of applying gates after the channel to transform to the diagonal basis, which is then calculated from a measured $n$-qubit probability distribution ${\rm Pr}(x)$. The probably distribution can be corrected for measurement error using the $\Gamma$ matrix (\ref{defGamma}), as explained above. Next, we use  (\ref{defL_1}) to simulate the ideal preparation of $\sigma'$. For a single qubit $i$, we obtain
\begin{eqnarray}
\begin{pmatrix} I \\ X \\ Y \\ Z \\ \end{pmatrix}
&=&
\begin{pmatrix} 
1 & 1 & 0 & 0 \\  
-1 & -1 & 2 & 0 \\ 
-1 & -1 & 0 & 2 \\ 
1 & -1 & 0 & 0 
\end{pmatrix} 
\begin{pmatrix} \pi_{0} \\ \pi_{1} \\ \pi_{+} \\ \pi_{+i} \\ \end{pmatrix}
= M \begin{pmatrix} \pi_{0} \\ \pi_{1} \\ \pi_{+} \\ \pi_{+i} \\ \end{pmatrix} 
\nonumber \\
&=&
M L_i^{-1} 
\begin{pmatrix} \rho_{0} \\ \rho_{1} \\ \rho_{+} \\ \rho_{+i} \\ \end{pmatrix},
\end{eqnarray}
relating ideal Pauli's to noisy initial states (\ref{nonidealStates}). Then similarly to (\ref{GammaEstCompact}), $\Phi_{\sigma \sigma'}$ is constructed from the measured values of $ {\rm tr} [\sigma \, \Phi( \rho_\lambda) ]$ for all $\sigma \in \lbrace I, X, Y Z \rbrace^n$ and $ {\lambda} \in \{ 0,1,+,+i \}^n$.

In conclusion, we have considered transition matrix error mitigation \cite{DewesPRL12,160304512,180411326,190411935,190708518,191001969,191113289,181010523,200614044,201008520,BialczakNatPhys10,NeeleyNat10,181102292} based on the matrix (\ref{defT}). By instead using the matrix $\Gamma$ defined in (\ref{defGamma}), we can correct measurement errors without contamination by state-preparation errors. We show how to obtain $\Gamma$ by performing single-qubit GST on each qubit of the register, plus $4^n$ additional measurements. This is an increase in sample complexity over the $2^n$ measurements required for a complete estimation of $T$, but the resulting error mitigation is rigorously exact, conditioned on the following assumptions:
\begin{enumerate}

\item The noisy classical states are separable.

\item The multiqubit POVM is diagonal in the classical basis. However the measurement operators are not assumed to be separable.

\item GST converges to a gateset that accurately models the input data, and the gauge choice accurately separates state preparation, gate, and measurement errors.

\end{enumerate}
Note that we don't assume that the $G_x$ and $G_y$ gates are truly single-qubit. Any entanglement with neighbors is seen by GST as another decoherence mechanism, which it will try to learn (see \cite{SI} for further discussion). 

Condition 1 means that we can learn the explicit form of the noisy initial states through single-qubit GST. However conditionally rigorous TMEM can be generalized to allow for entangled inputs, assuming they have local support: If a constant number $m$ qubits are entangled, then these entangled states can also be learned with $m$-qubit GST. Condition 2 is a sufficient  condition for the possibility of rigorous correction \cite{190708518,200201471}. 
Condition 3 means that the violation of the learned model by the data is negligible. GST may fail to converge if the noise is nonstationary or strongly non-Markovian. Condition 3 also reflects the impossibility for any tomographic technique to uniquely decompose observed nonidealities into separate state, gate, and POVM errors \cite{BlumeKohout13104492,181005631}. GST chooses a gauge to maximize the overall fit of the model to the ideal targets, and we are implicitly assuming that the resulting decomposition is reliable (our classical simulation \cite{SI} supports this). Conditioned on these assumptions, rigorous TMEM should help make high-precision and foundational quantum information science experiments possible with gate-based quantum computing platforms.

{\it Acknowledgments.} 
I am grateful to IBM Research for making their superconducting processors available to the quantum computing community. This work does not reflect the views or opinions of IBM or any of their employees. I would also like to thank the anonymous referees for suggestions that improved the presentation of this paper.

\bibliography{/Users/mgeller/Dropbox/bibliographies/algorithms,/Users/mgeller/Dropbox/bibliographies/applications,/Users/mgeller/Dropbox/bibliographies/books,/Users/mgeller/Dropbox/bibliographies/cm,/Users/mgeller/Dropbox/bibliographies/dwave,/Users/mgeller/Dropbox/bibliographies/general,/Users/mgeller/Dropbox/bibliographies/group,/Users/mgeller/Dropbox/bibliographies/ions,/Users/mgeller/Dropbox/bibliographies/links,/Users/mgeller/Dropbox/bibliographies/math,/Users/mgeller/Dropbox/bibliographies/ml,/Users/mgeller/Dropbox/bibliographies/nmr,/Users/mgeller/Dropbox/bibliographies/optics,/Users/mgeller/Dropbox/bibliographies/qec,/Users/mgeller/Dropbox/bibliographies/qft,/Users/mgeller/Dropbox/bibliographies/simulation,/Users/mgeller/Dropbox/bibliographies/software,/Users/mgeller/Dropbox/bibliographies/superconductors,/Users/mgeller/Dropbox/bibliographies/surfacecode,endnotes}




\clearpage

\setcounter{equation}{0}
\setcounter{figure}{0}
\setcounter{table}{0}
\setcounter{page}{1}
\setcounter{section}{0}
\setcounter{secnumdepth}{4}
\makeatletter
\renewcommand{\thesection}{\arabic{section}}
\renewcommand{\theequation}{S\arabic{equation}}
\renewcommand{\thefigure}{S\arabic{figure}}
\renewcommand{\bibnumfmt}[1]{[S#1]}
\clearpage
\onecolumngrid

\begin{center}
\Large{ Supplementary Information for \\ ``Conditionally rigorous mitigation of multiqubit measurement errors''}
\end{center}

\vspace{1cm}
\twocolumngrid

This document provides additional details about the validation and experimental implementation of the rigorous transition matrix error mitigation (TMEM) technique. In Sec.~\ref{Qubits} we describe the IBM Q online superconducting qubits used in the experiment and give calibration results (gate errors, coherence times, and single-qubit measurement  errors) provided by the backend. In Sec.~\ref{GST Primer} we provide a brief introduction to GST and discuss its limitations. In Sec.~\ref{GST Results} we give the detailed GST results. In Sec.~\ref{L Matrices} we provide the $L$ matrices used in the Mermin polynomial correction. In Sec.~\ref{Gamma Validation} we validate the technique used to calculate the $\Gamma$ matrices in the $n=1$ case by direct comparison with single-qubit GST. In Sec.~\ref{Mermin polynomials} we give the detailed Mermin polynomial measurement results and compare these with previous work. In Sec.~\ref{Simulation} we present the results of an end-to-end simulation of the correction of the Mermin polynomial $M_4$, further validating the rigorous TMEM technique.
 
\begin{widetext}

\section{Qubits}
\label{Qubits}

In this section we discuss the online superconducting qubits used in this work. Data was taken on the IBM Q processor ibmq\_santiago using the {BQP} software package developed by the author. BQP is a Python package developed to design, run, and analyze complex quantum computing and quantum information experiments using commercial backends. We demonstrate the rigorous TMEM technique using the qubits shown in Fig.~\ref{santiago chain figure}. Calibration data supplied by the backend is summarized in Table \ref{calibrationDataTable}. Here $T_{1,2}$ are the standard Markovian decoherence times, and
\begin{equation}
\epsilon = \frac{T(0|1) + T(1|0)}{2}
\end{equation}
is the single-qubit state-preparation and measurement (SPAM) error, averaged over initial classical states. The $U_2$ error column gives the single-qubit gate error measured by randomized benchmarking. The CNOT errors are also measured by randomized benchmarking.
 
\begin{table}[htb]
\centering
\caption{Calibration data provided by IBM Q for the ibmq\_santiago chip during the period of data acquisition.}
\begin{tabular}{c}
\begin{tabular}{|c|c|c|c|c|}
\hline
Qubit & $T_1 \ (\mu s)$ & $T_2  \ (\mu s)$ & SPAM error $\epsilon$ &  $U_2$ error \\
\hline 
$Q_0$ & 108.6  & 151.1 & 0.0139 & 2.98e-4  \\
$Q_1$ & 122.8  & 85.9 & 0.0137 &1.73e-4  \\
$Q_2$ & 120.1  & 92.4 & 0.0178 & 1.63e-4  \\
$Q_3$ & 157.7  & 90.3 & 0.0135 & 1.90e-4  \\
\hline 
\end{tabular}
\\
\begin{tabular}{|c|c|}
\hline 
CNOT gates &  CNOT error  \\
\hline 
\begin{tabular}{c|c} ${\rm CNOT}_{0,1}$ &  ${\rm CNOT}_{1,0}$  \\ \end{tabular} &   8.14e-3   \\
\hline 
\begin{tabular}{c|c} ${\rm CNOT}_{1,2}$ &  ${\rm CNOT}_{2,1}$  \\ \end{tabular} &   7.03e-3    \\
\hline 
\begin{tabular}{c|c} ${\rm CNOT}_{2,3}$ &  ${\rm CNOT}_{3,2}$  \\ \end{tabular} &   6.36e-3    \\
\hline 
\end{tabular} 
\\
\end{tabular}
\label{calibrationDataTable}
\end{table}

\clearpage

\begin{figure}
\includegraphics[width=7.0cm]{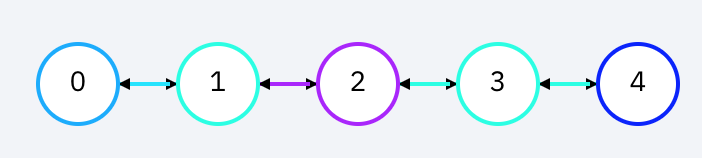} 
\caption{Layout of IBM Q device ibmq\_santiago. In this work we use qubits $Q_0$, $Q_1$, $Q_2$ and $Q_3$.}
\label{santiago chain figure}
\end{figure} 

\section{GST Primer}
\label{GST Primer}

Gate set tomography (GST) is a tomographic technique used to simultaneously estimate one or more initial states $\{\rho_i\}$, an informationally complete set of gates $\{ \Phi_i\}$, modeled here as CPTP superoperators $\rho \mapsto \rho' = \Phi(\rho)$, and a POVM $\{ E_x \}_{x \in \{0,1\}^n}$, all defined on a fixed register of $n$ qubits \cite{BlumeKohout13104492,MerkelPRA13,BlumeKohoutNat17,pygsti}. It works by measuring a large batch of circuits, designed to amplify errors by repeating carefully chosen germ sequences. After an iterative global optimization procedure the entire dataset is then fit to a global gateset model
\begin{eqnarray}
\big\lbrace \{\rho_i\}, \  \{ \Phi_i\} ,  \  \{ E_x \}  \big\rbrace.
\end{eqnarray}
 GST returns a representation of this gateset along with several fidelity measures and statistical confidence bounds.
 
For the purposes of conditionally rigorous TMEM, it is only necessary to estimate one initial state $\rho_0$ and two gates per qubit, namely $\pi/2$ rotations about $x$ and $y$, which we write a $G_x$ and $G_y$. On each qubit we measured 589 distinct circuits up to length 16 (18 including fiducials), and used the robust CPTP estimates from pyGSTi \cite{pygsti}. GST also produces a single-qubit POVM for each qubit, which we do not use here.

Like any tomographic technique, GST has limitations and can fail to produce an accurate estimate. There are two principle failure modes: First the gateset may fail to converge. The reason for this is that the data is fit to a {\it stationary} gateset, whereas real data contain non-Markovian noise and drift. GST provides global measures of the fit of data to the estimated model (see Fig.~\ref{modelViolation} below). In our experiment we obtained a ``good" fit to the estimated gateset. The second failure mode concerns the issue of tomography gauge. It is not possible for GST to produce a unique gateset; in particular, to uniquely separate a circuit error into state-preparation error, gate error, and POVM error. The most common approach, also followed here, is to choose this gauge to minimize the difference between the estimated gateset and the ideal one. 

We do not have to assume that $G_x$ and $G_y$ are truly single-qubit gates. Any entanglement with neighboring qubits is seen by GST as another decoherence mechanism, which it will learn as long as the associated noise is stationary.

To test the accuracy of GST in our experiment, we implemented a complete end-to-end simulation of the entire procedure. This is discussed below in Sec.~\ref{Simulation}. We find that GST is able to accurately learn a planted error model using the standard gauge choice. However we note that this gauge might not apply  to a highly asymmetric error model, where, for example, some elements of the model have much larger errors than others.

\section{GST Results}
\label{GST Results}

Here we give the detailed GST results provided by pyGSTi \cite{pygsti} for the $n \! = \! 4$ case. Tables \ref{gstSummaryTable} and \ref{gstErrorsTable} give the estimated single-qubit models and error measures, and Fig.~\ref{modelViolation} shows the model violation (caused by non-stationary noise, including drift) versus gate sequence length.

\begin{table}[htb]
\centering
\caption{Reconstructed $\rho_0$ state and two-outcome POVM estimated by GST for the IBM Q device ibmq\_santiago.}
\begin{tabular}{|c|c|c|c|c|}
\hline
Qubit & $\rho_{0}$ & $G_{\rm x}$ & $G_{\rm y}$ &  $E_0$ \\
\hline 
$Q_0$   
& $\begin{pmatrix} 0.9849 & 0.0539 \\ 0.0539 & 0.0151 \\ \end{pmatrix}$   
& $\begin{pmatrix} 1 & 0 & 0 & 0 \\ 0.0001 & 0.9999 & -0.0002 & -0.0001 \\ 0 & -0.0001 & -0.0001 & -1 \\ 0& 0.0001 & 0.9999 & -0.0001 \\ \end{pmatrix}$ 
& $\begin{pmatrix} 1 & 0 & 0 & 0 \\ 0 & 0.0002 & -0.0001 & 1 \\ 0 & -0.0001 & 1 & 0.0002 \\ 0 & -1 & -0.0002 & 0.0002 \\ \end{pmatrix}$ 
& $\begin{pmatrix} 0.9994 & -0.0005 \\ -0.0005 & 0.0080 \\ \end{pmatrix}$ \\
\hline
$Q_1$   
& $\begin{pmatrix} 0.9848 & 0.0534 \\ 0.0534 & 0.0152 \\ \end{pmatrix}$   
& $\begin{pmatrix} 1 & 0 & 0 & 0 \\ 0 & 1 & 0 & 0 \\ 0 & 0 & 0 & -1 \\ 0 & 0 & 1 & 0 \\ \end{pmatrix}$ 
& $\begin{pmatrix} 1 & 0 & 0 & 0 \\ 0 & -0.0002 & 0 & 1 \\ 0 & 0 & 1 & 0 \\ 0 & -1 & 0 & -0.0002 \\ \end{pmatrix}$ 
& $\begin{pmatrix} 0.9993 & 0 \\ 0 & 0.0082 \\ \end{pmatrix}$  \\
\hline
$Q_2$   
& $\begin{pmatrix} 0.9845 & 0.0537 \\ 0.0537 & 0.0155 \\ \end{pmatrix}$   
& $\begin{pmatrix} 1 & 0 & 0 & 0 \\ 0 & 1 & 0 & 0 \\ 0 & 0 & -0.0001 & -1 \\ 0 & 0 & 1 & -0.0001 \\ \end{pmatrix} $ 
& $\begin{pmatrix} 1& 0 & 0 & 0 \\ 0 & -0.0001 & 0 & 1\\ 0 & 0 & 1 & 0 \\ 0 & -1 & 0 & -0.0001 \\ \end{pmatrix}$ 
& $\begin{pmatrix} 0.9995 & -0.0010 \\ -0.0010 & 0.0086 \\ \end{pmatrix}$  \\
\hline
$Q_3$   
& $\begin{pmatrix} 0.9846 & 0.0540 \\ 0.0540 & 0.0154 \\ \end{pmatrix}$   
& $\begin{pmatrix} 1 & 0 & 0 & 0 \\ -0.0001 & 0.9997 & 0.0001 & 0.0003 \\ 0 & -0.0002 & -0.0003 & -0.9997 \\ 0.0001 & 0 & 0.9995 & -0.0002 \\ \end{pmatrix}$ 
& $\begin{pmatrix} 1 & 0 & 0 & 0 \\ 0.0001 & -0.0002 & 0.0001 & 0.9997 \\ 0 & -0.0001 & 0.9997 & 0 \\ 0 & -0.9998 & 0.0002 & 0 \\ \end{pmatrix}$ 
& $\begin{pmatrix} 1 & -0.0009 \\ -0.0009 & 0.0105 \\ \end{pmatrix}$  \\
\hline
\end{tabular}
\label{gstSummaryTable}
\end{table}

\begin{table}[htb]
\centering
\caption{Gate error measures.}
\begin{tabular}{|c|c|c|c|c|}
\hline
Qubit & Gate & Entanglement infidelity & Trace distance/2 & Eigenvalue entanglement infidelity \\
\hline 
$Q_0$& $G_{\rm x}$ & 6.31e-5  & 1.47e-4  & 6.31e-5   \\
           & $G_{\rm y}$ & 1.46e-5  & 1.34e-4  & 1.46e-5  \\
\hline 
$Q_1$& $G_{\rm x}$ & 3.84e-10  & 1.95e-5  & 5.88e-15   \\
           & $G_{\rm y}$ & 1.14e-8  & 1.07e-4  & 2.09e-14  \\
\hline 
$Q_2$& $G_{\rm x}$ & 1.08e-9  & 3.28e-5  & 4.21e-15   \\
           & $G_{\rm y}$ & 4.98e-9  & 7.06e-5  & 2.22e-16  \\
\hline 
$Q_3$& $G_{\rm x}$ & 2.74e-4  & 3.23e-4  & 2.74e-4   \\
           & $G_{\rm y}$ & 2.00e-4  & 2.20e-4  & 2.00e-4  \\
\hline
\end{tabular}
\label{gstErrorsTable}
\end{table}

\clearpage
\begin{figure}
\includegraphics[width=10.0cm]{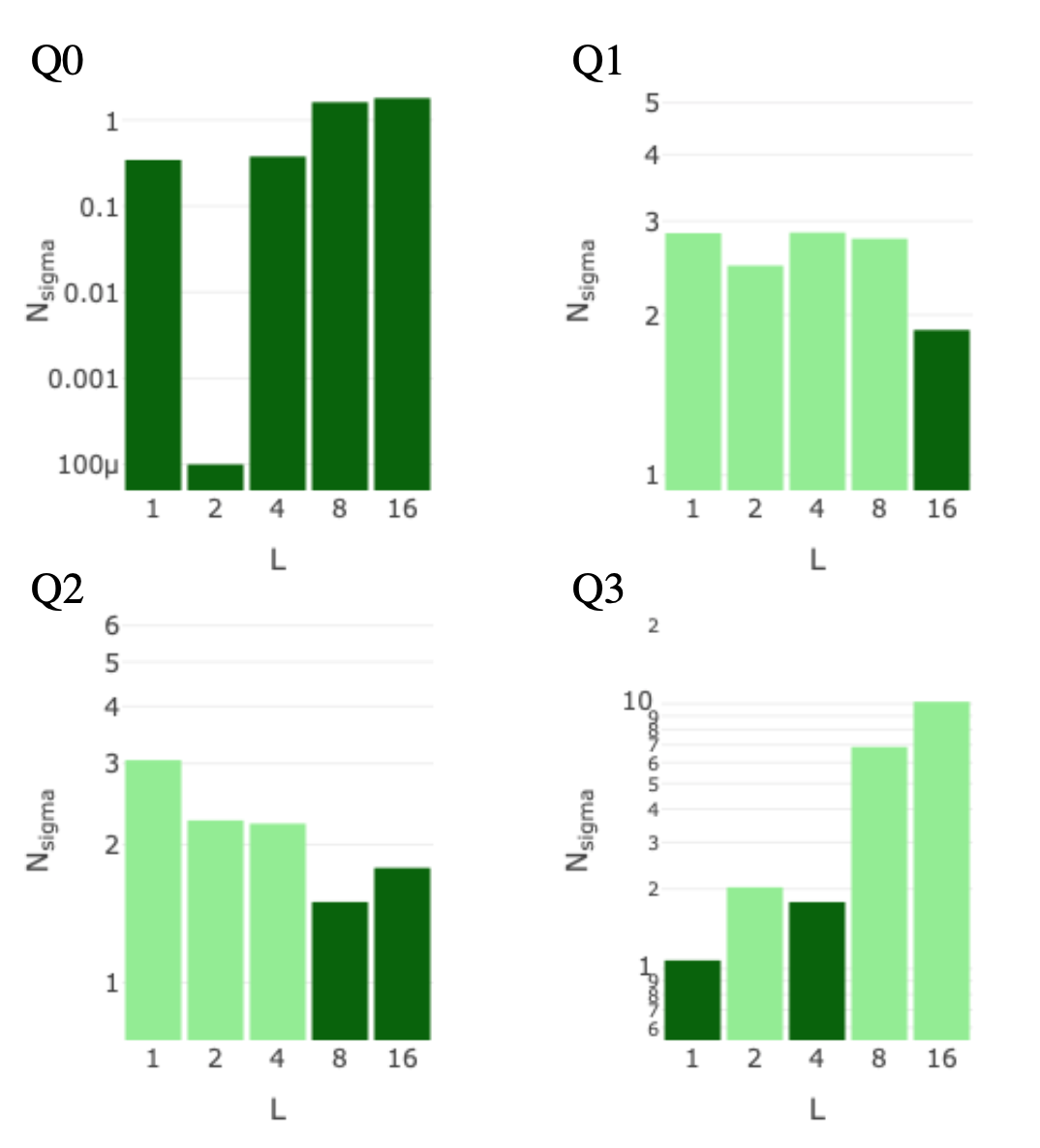} 
\caption{GST model violation versus sequence length {\sf L}.}
\label{modelViolation}
\end{figure} 

\section{${\bm L}$ Matrices}
\label{L Matrices}

In Table \ref{LTable} we provide the calculated $L$ matrices for the qubits used in the experiments for the $n \! = \! 4$ case.

\begin{table}[htb]
\centering
\caption{$L$ matrices.}
\begin{tabular}{|c|c|}
\hline
Qubit& $L$  \\
\hline 
$Q_0$  & $\begin{pmatrix} 0.9164 & -0.0535 & 0.1079 & 0.0292 \\ -0.0243 & 0.9454 & 0.1080 & -0.0291 \\ -0.0535 & 0.0543 & 0.9699 & 0.0293 \\ -0.0534 & -0.0244 & 0.1082 & 0.9697 \\ \end{pmatrix}$   \\
\hline
$Q_1$  & $\begin{pmatrix} 0.9170 & -0.0527 & 0.1067 & 0.0289 \\ -0.0238 & 0.9459 & 0.1067 & -0.0288 \\ -0.0527 & 0.0542 & 0.9696 & 0.0289 \\ -0.0526 & -0.0238 & 0.1067 & 0.9697 \\ \end{pmatrix}$   \\
\hline
$Q_2$  & $\begin{pmatrix} 0.9157 & -0.0533 & 0.1074 & 0.0302 \\ -0.0231 & 0.9458 & 0.1074 & -0.0301 \\ -0.0533 & 0.0542 & 0.9689 & 0.0302 \\ -0.0532 & -0.0231 & 0.1073 & 0.9689 \\ \end{pmatrix}$   \\
\hline
$Q_3$  & $\begin{pmatrix} 0.9153 & -0.0538 & 0.1081 & 0.0304 \\ -0.0232 & 0.9451 & 0.1081 & -0.0301 \\ -0.0537 & 0.0544 & 0.9690 & 0.0304 \\ -0.0527 & -0.0230 & 0.1077 & 0.9680 \\ \end{pmatrix}$   \\
\hline
\end{tabular}
\label{LTable}
\end{table}

\section{${\bm \Gamma}$ Validation}
\label{Gamma Validation}

Here we validate the $\Gamma$ estimation technique, based on quasiprobability decomposition, for the special case of a 1-qubit register, by comparing it with that obtained from the POVM directly estimated by single-qubit GST \cite{200201471}, which we denote by $\Gamma_{\rm POVM}$. The results are shown in Table \ref{gammaValidationTable}. The concurrently measured single-qubit $T$ matrices are also given, as are the Frobenius distances
$\| \Gamma-T \|_{\rm F}$ and $\| \Gamma-\Gamma_{\rm POVM} \|_{\rm F}$.

\begin{table}[htb]
\centering
\caption{$\Gamma$ versus  $\Gamma_{\rm POVM}$ for the case $n\!=\!1$.}
\begin{tabular}{|c|c|c|c|c|c|}
\hline
Qubit& $\Gamma$ &   $\Gamma_{\rm POVM}$  & $T$ & $\| \Gamma-T \|_{\rm F}$ & $\| \Gamma-\Gamma_{\rm POVM} \|_{\rm F}$ \\
\hline 
$Q_0$  & $\begin{pmatrix} 0.9981 & 0.0063 \\ 0.0019 & 0.9937 \\ \end{pmatrix}$ & $\begin{pmatrix} 0.9994 & 0.0080 \\ 0.0006 & 0.9920 \\ \end{pmatrix}$ & $\begin{pmatrix} 0.9838 & 0.0257 \\ 0.0163 & 0.9742 \\ \end{pmatrix}$ &3.42e-2 & 3.11e-3 \\
\hline 
$Q_1$  & $\begin{pmatrix} 0.9992 & 0.0060 \\ 0.0008 & 0.9940 \\ \end{pmatrix}$ & $\begin{pmatrix} 0.9993 & 0.0082 \\ 0.0007 & 0.9918 \\ \end{pmatrix}$ & $\begin{pmatrix} 0.9845 & 0.0243 \\ 0.0155 & 0.9758 \\ \end{pmatrix}$ & 3.32e-2 & 3.18e-3 \\
\hline 
$Q_2$  & $\begin{pmatrix} 0.9977 & 0.0108 \\ 0.0023 & 0.9892 \\ \end{pmatrix}$ & $\begin{pmatrix} 0.9995 & 0.0086 \\ 0.0005 & 0.9914 \\ \end{pmatrix}$ & $\begin{pmatrix} 0.9819 & 0.0294 \\ 0.0181 & 0.9706 \\ \end{pmatrix}$ &3.45e-2 &3.99e-3 \\
\hline 
$Q_3$  & $\begin{pmatrix} 0.9986 & 0.0125 \\ 0.0014 & 0.9875 \\ \end{pmatrix}$ & $\begin{pmatrix} 1.0000 & 0.0105 \\ 0.0000 & 0.9895 \\ \end{pmatrix}$ & $\begin{pmatrix} 0.9829 & 0.0253 \\ 0.0171 & 0.9748 \\ \end{pmatrix}$ & 2.85e-2 & 3.53e-3 \\
\hline 
\end{tabular}
\label{gammaValidationTable}
\end{table}

\clearpage

\section{Mermin polynomial data}
\label{Mermin polynomials}

In this section the results of the $N\!=\!16$ independent measurements of the Mermin polynomials $M_n$, for $n=3$ and 4, are presented as histograms. For each $M_n$, the raw data values, the values corrected by a concurrently measured $n$-qubit $T$ matrix, and the values corrected by the $\Gamma$ matrix (using the $L$ matrices from Table \ref{LTable}) are shown separately, but plotted on the same scale. Figures~\ref{M3_Raw}, \ref{M3_T}, and \ref{M3_Gamma} contain the $M_3$ results, and Figs.~\ref{M4_Raw}, \ref{M4_T}, and \ref{M4_Gamma} contain the $M_4$ results. 

\begin{figure}
\includegraphics[width=12.0cm]{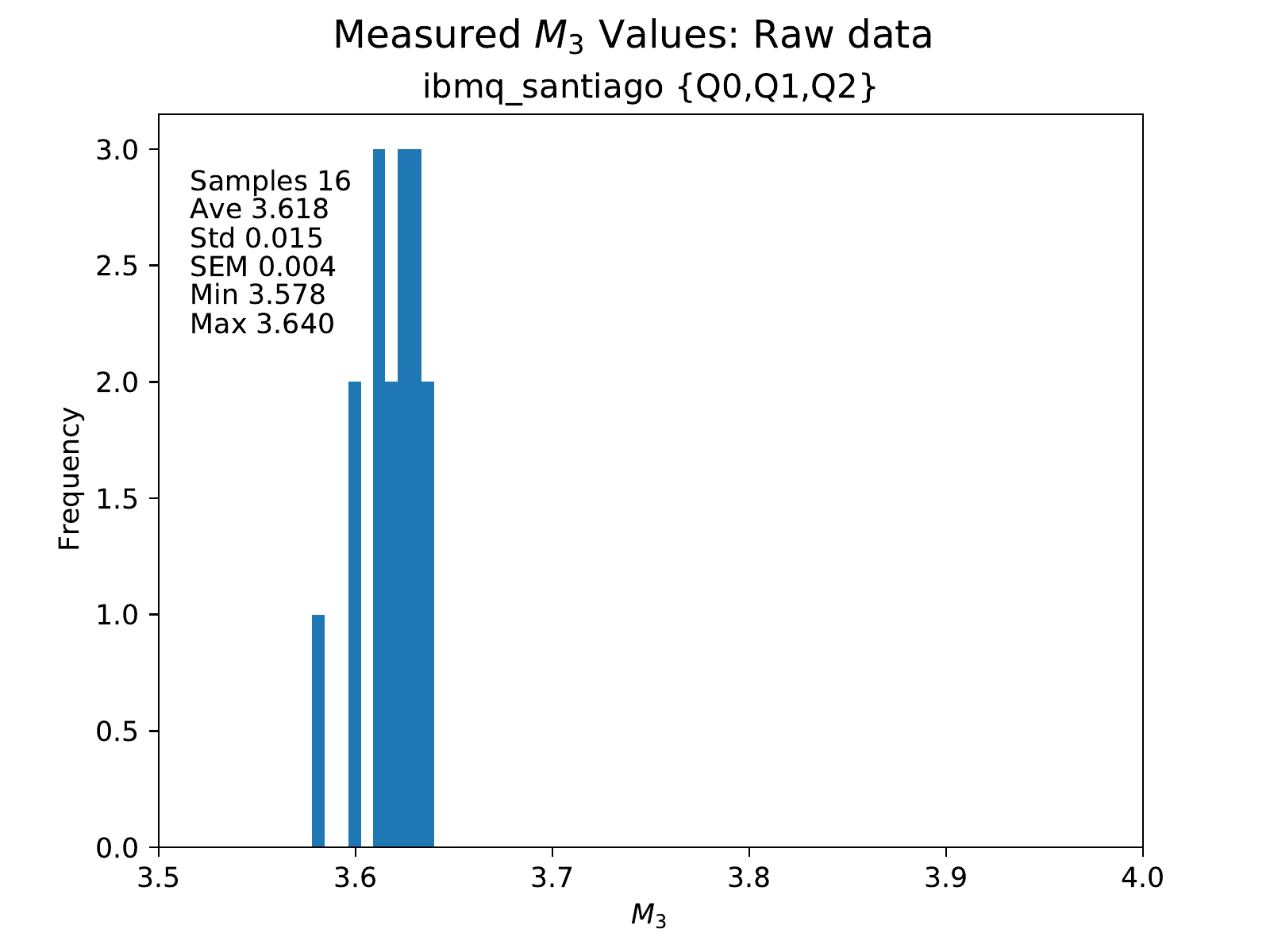} 
\caption{}
\label{M3_Raw}
\end{figure} 

\begin{figure}
\includegraphics[width=12.0cm]{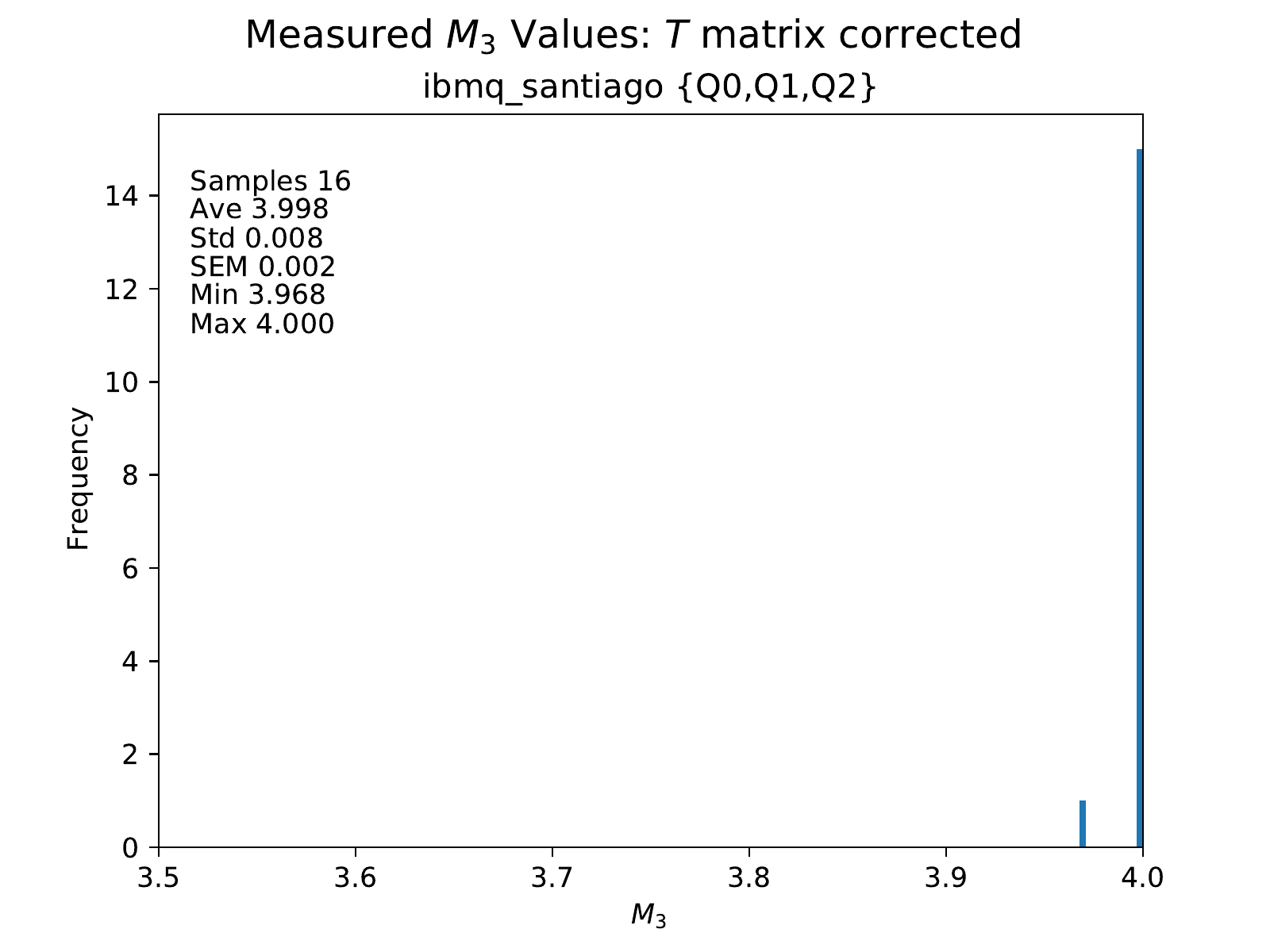} 
\caption{}
\label{M3_T}
\end{figure} 

\begin{figure}
\includegraphics[width=12.0cm]{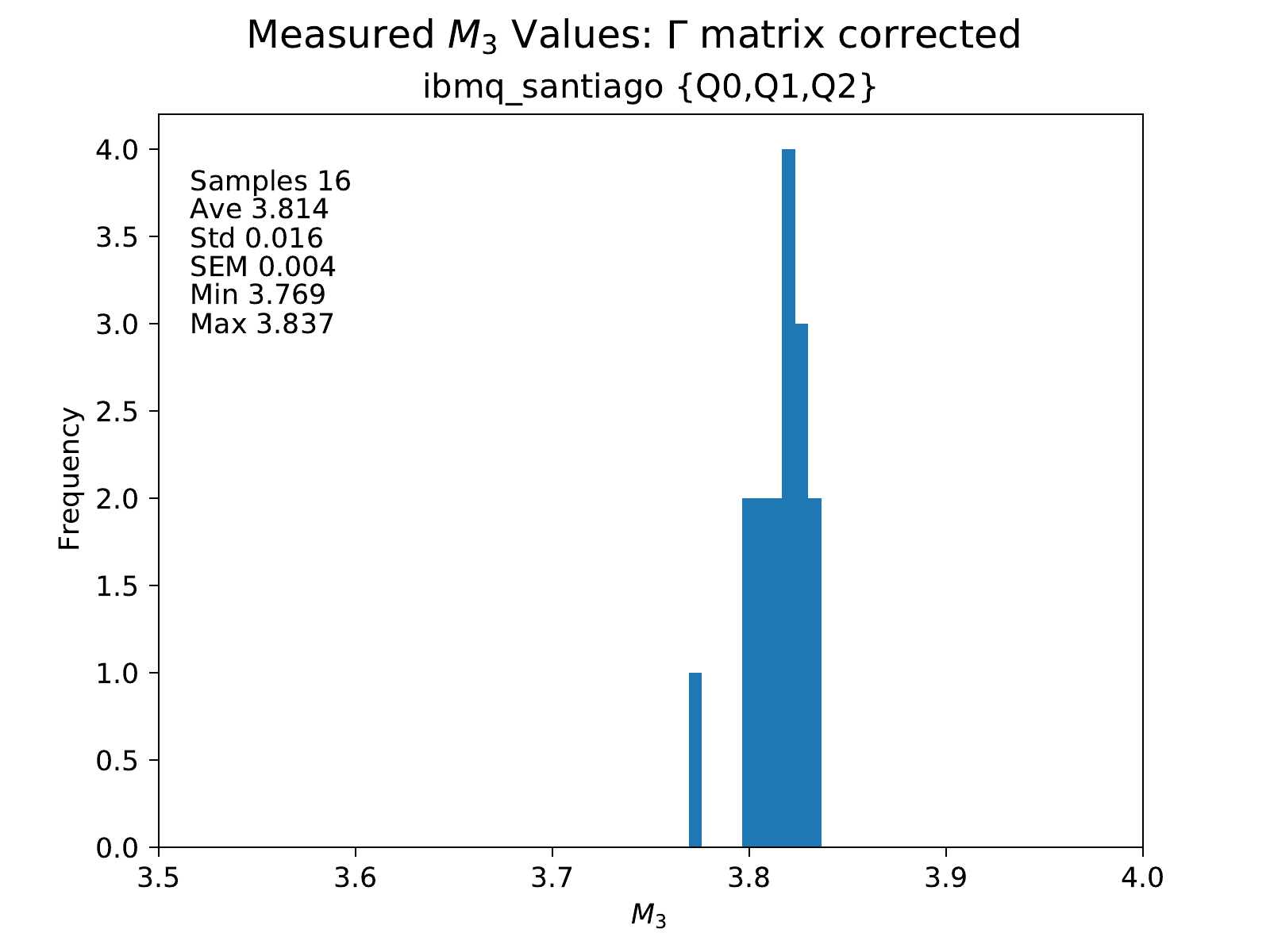} 
\caption{}
\label{M3_Gamma}
\end{figure} 

\begin{figure}
\includegraphics[width=12.0cm]{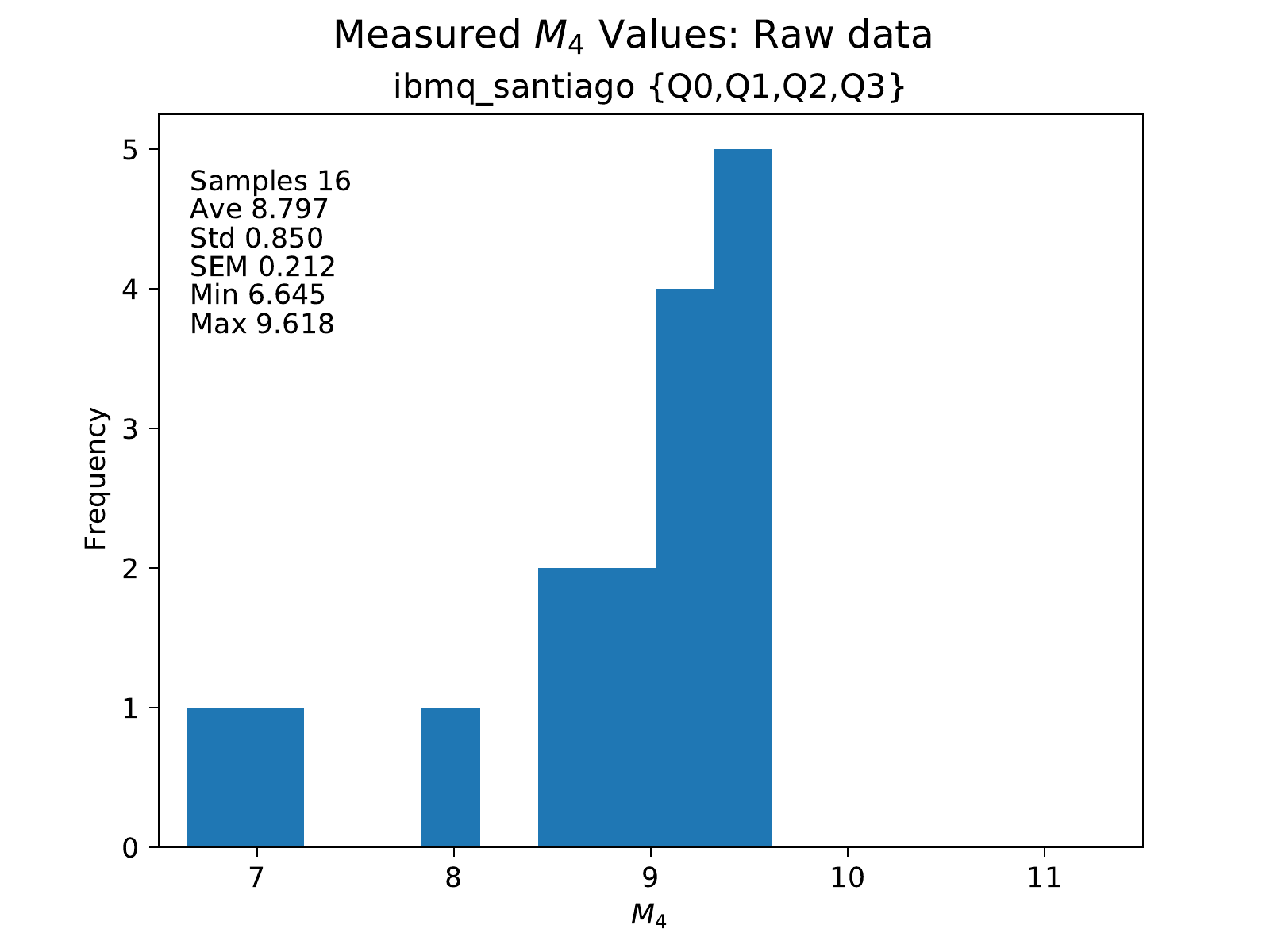} 
\caption{}
\label{M4_Raw}
\end{figure} 

\begin{figure}
\includegraphics[width=12.0cm]{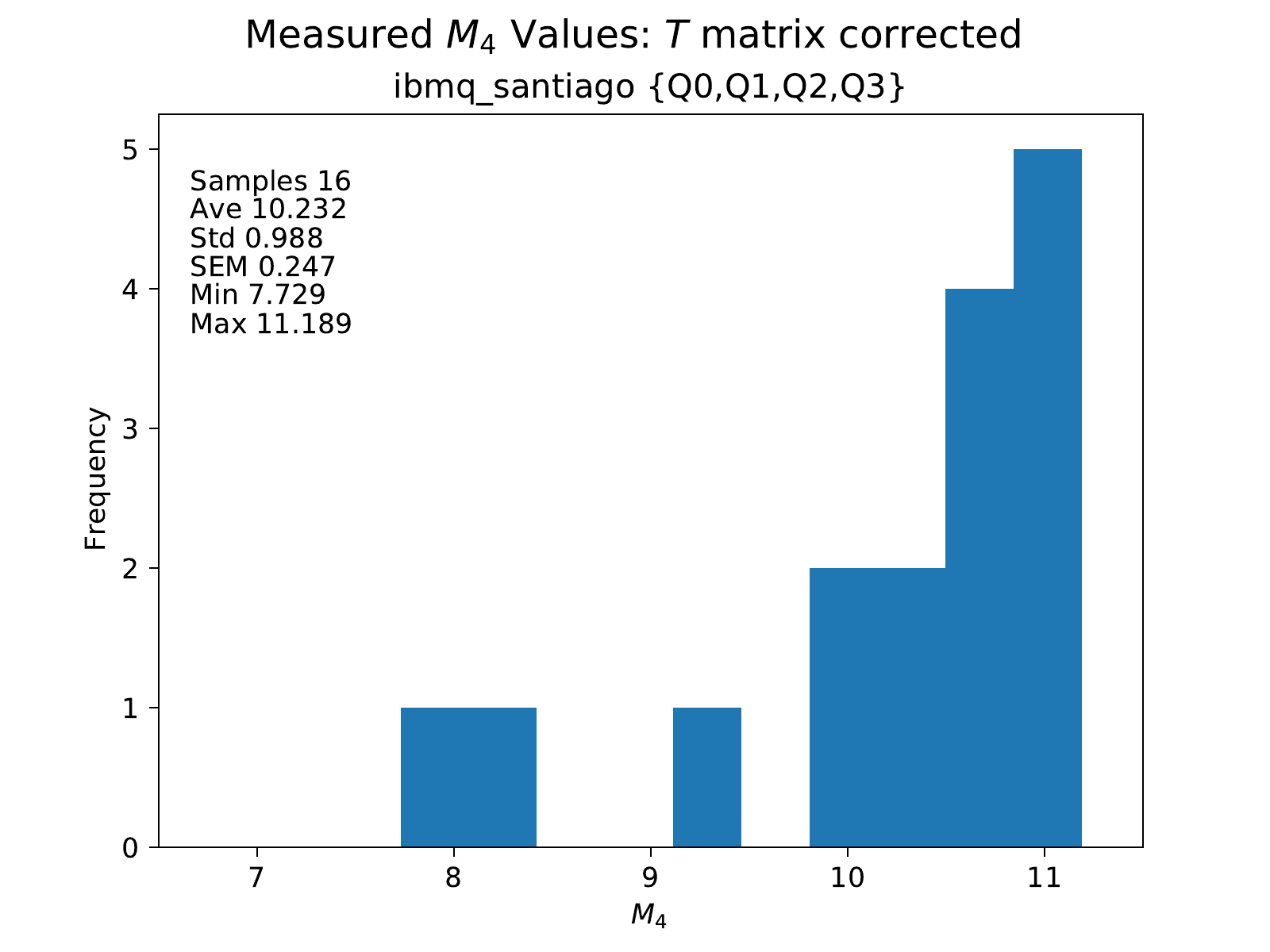} 
\caption{}
\label{M4_T}
\end{figure} 

\begin{figure}
\includegraphics[width=12.0cm]{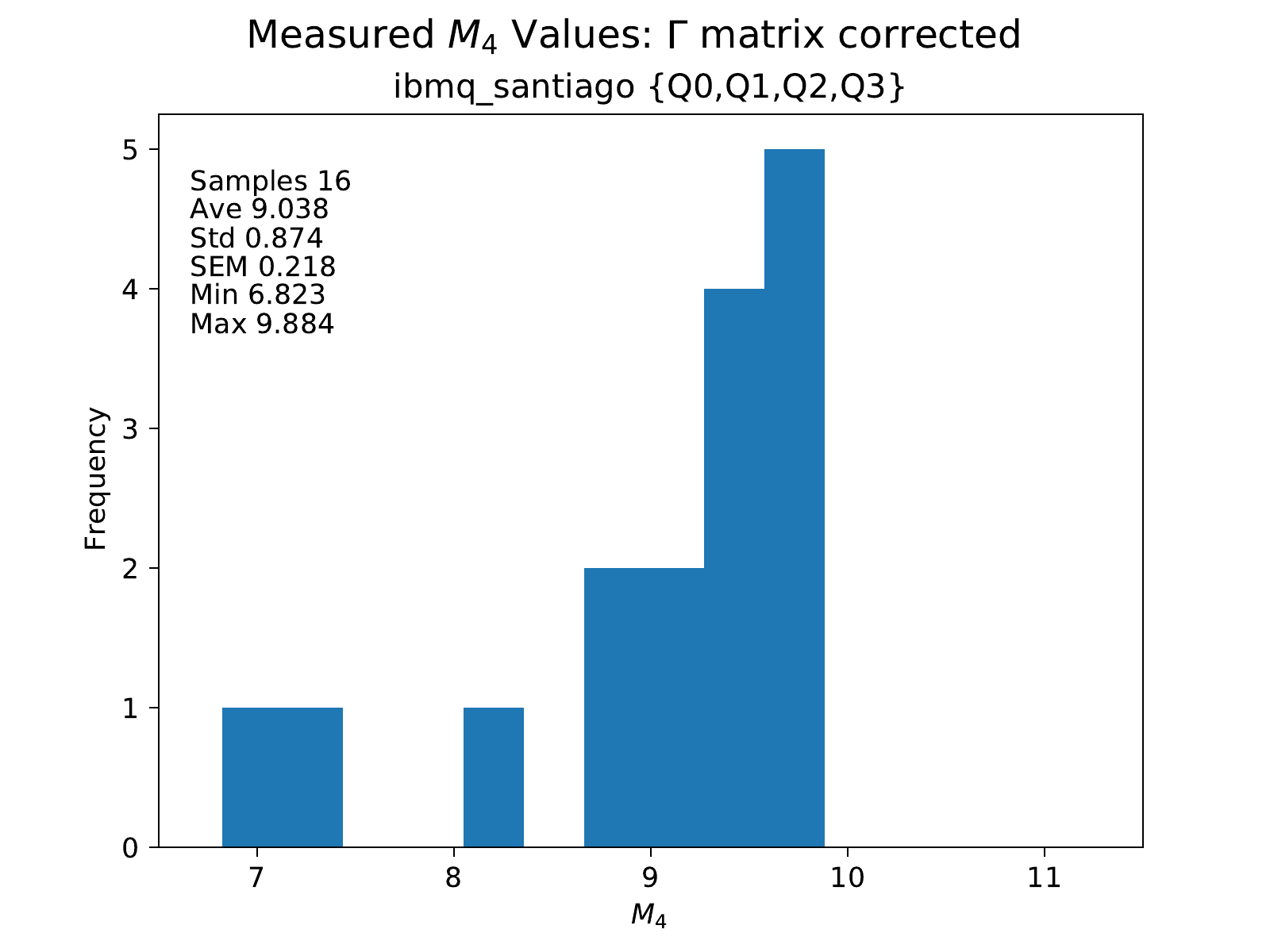} 
\caption{}
\label{M4_Gamma}
\end{figure} 

\clearpage

There is a long history of measuring Mermin polynomials on superconducting qubits.
In Table \ref{merminComparisonTable} we compare our measured values of $M_3$ and $M_4$ with results obtained previously.
 
\begin{table}[htb]
\centering
\caption{Comparison of Mermin polynomial measurements. Here $n$ is the register size, LR is the maximum value consistent with local realism, and QM is the maximum value allowed by quantum mechanics. The last columns gives our error-mitigated results. 
}
\begin{tabular}{|c|c|c|c|c|c|c|c|c|}
\hline
$n$ & LR & QM & Neeley {\it et al.}~\cite{NeeleyNat10} & DiCarlo {\it et al.}~\cite{DiCarloNat10} & Alsina \& Latorre~\cite{AlsinaPRA16} & Garcia {\it et al.}~\cite{171205642}  & Gonzales {\it et al.}~\cite{200511271} & {\bf This work} \\
\hline
3 &2 & 4 &  2.08 & 3.40 & 2.85 & 2.84 & 3.34 & {\bf 3.81}  \\
\hline 
4 & 4 & $8\sqrt{2}$ &&& 4.81 & 5.42 & 9.07  & {\bf 9.04} \\
\hline 
\end{tabular}
\label{merminComparisonTable}
\end{table}

\section{Simulation}
\label{Simulation}

In this section we present the results of a complete end-to-end simulation of the conditionally rigorous TMEM technique applied the estimation of the 4-qubit Mermin polynomial 
\begin{eqnarray}
&M_4 = \langle XXXY \rangle + \langle XXYX \rangle + \langle XYXX \rangle +  \langle YXXX \rangle + \langle XXYY \rangle + \langle XYXY \rangle + \langle XYYX \rangle +  \langle YXXY \rangle &\nonumber \\
&+ \langle YXYX \rangle + \langle YYXX \rangle - \langle XXXX \rangle -  \langle XYYY \rangle - \langle YXYY \rangle - \langle YYXY \rangle - \langle YYYX \rangle -  \langle YYYY \rangle \ \ &
\end{eqnarray}
on the depolarized GHZ state
\begin{equation}
\rho = \big(1-\eta \big) \, |\psi \rangle \langle \psi | + \eta \, \frac{I}{d},
\label{defRhoNoisy}
\end{equation}
where $\eta \in [0,1]$ is a noise parameter,
\begin{equation}
|\psi \rangle = \frac{ |0000\rangle + e^{3 \pi i /4}  |1111\rangle}{\sqrt 2},
\end{equation}
and $d=16$. We do this for all $0 \le \eta \le 1$. The simulation consists of the following steps:
\begin{enumerate}

\item[A.] First construct a simplified but realistic error model for a noisy 4-qubit register, including imperfect $|0\rangle$ state preparation and $\pi/2$ rotations $G_{\rm x,y}$, as well as a noisy multiqubit POVM. The fidelities of the $|0\rangle$ states and gates are chosen to approximately correspond to that measured on the ibmq\_santago chip. The noisy 16-element POVM is diagonal but otherwise random, with the noise strength again consistent with ibmq\_santago. Importantly, the POVM elements are {\it not} given as tensor products of single-qubit measurement operators. Call this the {\it exact error model} to contrast it with the approximate one predicted by GST in step B. Note that while the noisy states and gates are homogeneous across the register, the measurement errors are not.

\item[B.] Perform single-qubit GST on each qubit in the register using synthetic input data generated by the exact error model, to approximately reconstruct the noisy $|0\rangle$ state and $\pi/2$ gates. Single-qubit measurement operators are also produced by GST but are not used here. Call this the {\it approximate error model} and denote it by $\rho_0^{\rm est}$, $G_{\rm x,y}^{\rm est}$, and $E_0^{\rm est}$ for each qubit.

\item[C.] Compute the $L$ matrices from the GST output of step B.

\item[D.] Use the exact error model from step A to simulate the experimental measurement of the $\Gamma$ matrix. For comparison, also simulate the measurement of the $T$ matrix. Call these the {\it synthetic} $\Gamma$ and $T$ matrices.

\item[E.] Finally, assume that the noisy GHZ state (\ref{defRhoNoisy}) has been prepared in the register, and simulate the measurement of the 4-qubit Mermin polynomial (\ref{defM4}) in the presence of the noisy multiqubit POVM from step A. The depolarization in (\ref{defRhoNoisy}) is intended to account for all errors incurred during the estimation of $M_4$, including decoherence and imperfect tomography, but excluding multiqubit measurement errors (which are included separately in the noisy POVM). Therefore we can calculate both the noisy and exact measurement-error-free values of $M_4$. Then perform TMEM on the Mermin polynomial with the synthetic $\Gamma$ and $T$ matrices, and compare with the exact values.

\end{enumerate}

\clearpage

The simulation is repeated and averaged over $N\!=\! 16$ random POVM samples. Probabilities are assumed to be estimated with a large number of measurement samples and we do not include sampling errors in the simulation. Therefore, the only obstacle preventing perfect recovery of the measurement-error-free value of $M_4$ is the accuracy of the error model estimated by GST. In particular, if step B of the simulation is bypassed, and GST is rigged to return the exact error model of step A, exact values of the Mermin polynomial are always obtained. 

Steps A-E are explained in detail for the interested reader in the following subsections. Here we briefly summarize the main result of the simulation, given in Table~\ref{mermin simulation table}. Here LR is the maximum value consistent with local realism. QM is the maximum value allowed by quantum mechanics. Simulated data are reported in the remaining columns: Exact$(\eta \!=\! 0.2)$ is the Mermin polynomial (\ref{defM4}) calculated for the state (\ref{defRhoNoisy}) with $\eta = 0.2$, chosen for consistency with the real experiment on ibmq\_santago. Raw data is the same as Exact$(\eta \!=\! 0.2)$, but now calculated with the noisy multiqubit POVM from step A. The $T$ and $\Gamma$ columns give the polynomial values after correcting the synthetic probabilities with the synthetic $T$ and $\Gamma$ matrices. We observe that the average TMEM value based on $\Gamma$ is close to the exact value (they become equal in the limit that GST works perfectly), validating the rigorous technique. However, we speculate that GST might need to choose its gauge differently on highly {\it unbalanced} error models, where one error source is much larger or smaller than the others. 

\begin{table}[htb]
\centering
\caption{Simulated measurement and mitigation of the 4-qubit Mermin polynomial. Simulation results are reported as the average over $N\!=\! 16$ random POVM samples plus/minus the standard error $\sigma / \sqrt{N}$, where $\sigma^2$ is the variance of the $N$ samples. 
}
\begin{tabular}{|c|c|c|c|c|c|c|}
\hline
 & LR & QM & Exact$(\eta \!=\! 0.2)$ & Raw data & $T$ matrix & $\Gamma$ matrix \\
\hline
$M_4$ & 4 & $8\sqrt{2}$ & 9.051 & 8.962 $\pm$ 1.8e-3   & 9.821  $\pm$ 2.0e-5 & 9.067 $\pm$  9.2e-3  \\
\hline 
\end{tabular}
\end{table}

\subsection{Error model}

First we build a simplified model for the noisy $|0\rangle$ states,  $\pi/2$ rotations, and measurement operators consistent with those estimated by GST on the ibmq\_santoago chip. The model is simplified in that noise in $|0\rangle$ and the gates is modeled by depolarizing noise. The prepared $|0\rangle$ states on each qubit are given by
\begin{equation}
\rho_0 = \begin{pmatrix} 0.99 & 0 \\ 0 & 0.01 \\ \end{pmatrix},
\end{equation}
which results from a 2e-2 depolarization error applied to the ideal $|0\rangle$ state. The state fidelity $F = \langle 0 | \rho_0 | 0 \rangle$ is 0.99, consistent with the experimental values reported by GST. In the Pauli basis this becomes
\begin{equation}
|\rho_0\rangle \rangle = \begin{pmatrix} 0.50 \\ 0 \\ 0 \\ 0.49 \end{pmatrix},
\end{equation}
where the components of any single-qubit state $\rho$ are given by $|\rho \rangle \rangle_\sigma = {\rm Tr}(\rho \sigma)/2$ for  $\sigma \in \{ I, X, Y, Z \}$.

The gates $G_{\rm x,y}$ are similarly modeled as superoperators ${\cal G}_\mu$ corresponding to ideal $\pi/2$ rotations $e^{-i \pi X/4}$ and $e^{-i \pi Y/4}$ followed by a depolarization channel with 2e-4 depolarization error. In the Pauli basis these are $4 \! \times \! 4$ matrices with elements $(G_\mu)_{\sigma \sigma'} =  {\rm Tr} \, [\sigma \,{\cal G}_\mu(\sigma') ] /2$, where $\mu = {\rm x}, {\rm y}$. In the exact error model they are given by
\begin{equation}
G_{\rm x} = 
\begin{pmatrix}
1 & 0 & 0 & 0 \\
0 & 0.9998 & 0 & 0 \\
0 & 0 & 0 & -0.9998  \\
0 & 0 & 0.9998 & 0 \\
\end{pmatrix}
\end{equation}
and
\begin{equation}
G_{\rm y} = 
\begin{pmatrix}
1 & 0 & 0 & 0 \\
0 & 0 & 0 & 0.9998 \\
0 & 0 & 0.9998 & 0 \\
0 & -0.9998 & 0 & 0 \\
\end{pmatrix}
\end{equation}
on each qubit. The gate fidelities $F = [{\rm Tr} (G_{\mu, {\rm t}}^\dagger G_\mu) \! + \! 2]/6$ with the unitary targets $G_{\mu, {\rm t}}$ are 0.9999, consistent with the experimental values reported by GST.

The noisy POVM was generated iteratively by starting with the set of ideal projectors and randomly introducing small diagonal crosstalk errors in a way that guarantees that the 16 matrices remain positive semidefinite and that the set is properly normalized. The process is repeated until the noise level in the resulting $\Gamma$ matrix (measured by the Frobenius distance from the identity, $\| \Gamma-I \|_{\rm F} $) reaches the desired value. The synthetic POVM is not separable and contains correlated multiqubit measurement errors. The state $\rho_0$, gates $G_{\rm x,y}$, and POVM constitute the exact model error. The model does not include CNOT gates, which are not needed because the GHZ state preparation is described by the separate error model (\ref{defRhoNoisy}).

\subsection{GST}

Next we use pyGSTi \cite{pygsti} to run GST on synthetic data generated by the exact error model (using the same hyperparameter settings as used in the real experiment). The output of GST is the {\it approximate error model}. As explained above, the approximate error model includes a {\it separable} estimate of the POVM (tensor products of the $E_0^{\rm est}$ and $E_1^{\rm est} = I - E_0^{\rm est}$) , which we do not use. The approximate error model for the first of 16 random POVM instances is given in Table \ref{approximate error model table}. The measurement operators $E_0^{\rm est}$ differ from qubit to qubit because the noisy POVM from step A is not homogeneous across the qubits. While the gates are correctly recovered to high accuracy, there is a small variation in the reconstructed initial states $\rho_{0}^{\rm est} $ due to a small amount of measurement error bleeding into state preparation error.

\begin{table}[htb]
\centering
\caption{Approximate error model estimated by GST using synthetic input data generated by the exact error model.}
\begin{tabular}{|c|c|c|c|c|}
\hline
Qubit & $\rho_{0}^{\rm est} $ & $G_{\rm x}^{\rm est}$ & $G_{\rm y}^{\rm est}$ &  $E_0^{\rm est}$ \\
\hline 
$1$   
& $\begin{pmatrix} 0.9915 & 0 \\  0 & 0.0085 \\ \end{pmatrix}$   
& $\begin{pmatrix} 1 & 0 & 0 & 0 \\  0 & 0.9998 & 0 & 0 \\ 0 & 0 & 0 & -0.9997 \\ 0 & 0 & 0.9998 & 0 \\ \end{pmatrix}$ 
& $\begin{pmatrix} 1 & -0 & 0 & 0 \\ 0 & 0 & 0 & 0.9997 \\ 0 & 0 & 0.9998 & 0 \\ 0 & -0.9998 & 0 & 0 \\ \end{pmatrix}$ 
& $\begin{pmatrix} 0.9937 & 0 \\ 0 & 0.0023 \\ \end{pmatrix}$ \\
\hline
$2$   
& $\begin{pmatrix} 0.9895 & 0 \\ 0 & 0.0105 \\ \end{pmatrix}$   
& $\begin{pmatrix} 1 & 0 & 0 & 0 \\ 0 & 0.9998 & 0 & 0 \\ 0 & 0 & 0 & -0.9997 \\ 0 & 0 & 0.9998 & -0 \\ \end{pmatrix} $ 
& $\begin{pmatrix} 1 & 0 & 0 & 0 \\ 0 & 0 & 0 & 0.9997 \\ 0 & 0 & 0.9998 & 0 \\ 0 & -0.9998 & 0 & -0 \\ \end{pmatrix}$ 
& $\begin{pmatrix} 0.9937 & 0 \\ 0 & 0.0042 \\ \end{pmatrix}$  \\
\hline
$3$   
& $\begin{pmatrix} 0.9873 & 0 \\ 0 & 0.0127 \\ \end{pmatrix} $   
& $\begin{pmatrix} 1 & 0 & 0 & 0 \\ 0 & 0.9998 & 0 & 0 \\ 0 & 0 & 0 & -0.9997 \\ 0 & 0 & 0.9998 & 0 \\ \end{pmatrix}$ 
& $\begin{pmatrix} 1 & 0 & 0 & 0 \\ 0 & 0 & 0 & 0.9997 \\ 0 & 0 & 0.9998 & 0 \\ 0 & -0.9998 & 0 & 0 \\ \end{pmatrix}$ 
& $\begin{pmatrix} 0.9958 & 0 \\ 0 & 0.0085 \\ \end{pmatrix}$  \\
\hline
$4$   
& $\begin{pmatrix} 0.9906 & 0 \\  0 & 0.0094 \\ \end{pmatrix}$   
& $\begin{pmatrix} 1 & 0 & 0 & 0 \\ 0 & 0.9998 & 0 & 0 \\ 0 & 0 & 0 & -0.9997 \\ 0 & 0 & 0.9998 & 0 \\ \end{pmatrix}$ 
& $\begin{pmatrix} 1 & 0 & 0 & 0 \\ 0 & 0 & 0 & 0.9997 \\ 0 & 0 & 0.9998 & 0 \\ 0 & -0.9998 & 0 & 0 \\ \end{pmatrix}$ 
& $\begin{pmatrix} 0.9958 & 0 \\ 0 & 0.0053 \\ \end{pmatrix}$  \\
\hline
\end{tabular}
\label{approximate error model table}
\end{table}

\subsection{$L$ matrices}

The $L$ matrix calculated from the exact error model (the same on every qubit) is
\begin{equation}
L = 
\begin{pmatrix} 0.9900 & 0.0100 & 0 & 0 \\ 0.0102 & 0.9898 & 0 & 0 \\ 0.0101 & 0.0101 & 0.9798 & 0 \\ 0.0103 & 0.0103 & 0 & 0.9794 \\ \end{pmatrix}.
\label{L synthetic}
\end{equation}
The $L$ matrices calculated from synthetic GST data are all close to (\ref{L synthetic}) and are given in Table~\ref{simulated L table} for the first of 16 random POVM instances.

\clearpage

\begin{table}[htb]
\centering
\caption{$L$ matrices calculated from the approximate error model produced by GST.}
\begin{tabular}{|c|c|}
\hline
Qubit& $L$  \\
\hline 
$1$  & $\begin{pmatrix}
0.9915 & 0.0085 & 0 & 0 \\ 0.0087 & 0.9913 & 0 & 0 \\ 0.0086 & 0.0087 & 0.9827 & 0 \\ 0.0089 & 0.0089 & 0 & 0.9823 \\ \end{pmatrix}$   \\
\hline
$2$  & $\begin{pmatrix} 0.9895 & 0.0105 & 0 & 0 \\ 0.0108 & 0.9893 & 0 & 0 \\ 0.0106 & 0.0107 & 0.9787 & 0 \\ 0.0109 & 0.0109 & 0 & 0.9782 \\ \end{pmatrix} $   \\
\hline
$3$  & $\begin{pmatrix} 0.9874 & 0.0127 & 0 & 0 \\ 0.0129 & 0.9871 & 0 & 0 \\ 0.0128 & 0.0128 & 0.9744 & 0 \\ 0.0131 & 0.0131 & 0 & 0.9739 \\ \end{pmatrix}$   \\
\hline
$4$  & $\begin{pmatrix} 0.9906 & 0.0095 & 0 & 0 \\ 0.0097 & 0.9903 & 0 & 0 \\ 0.0096 & 0.0096 & 0.9808 & 0 \\ 0.0099 & 0.0098 & 0 & 0.9803 \\ \end{pmatrix}$   \\
\hline
\end{tabular}
\label{simulated L table}
\end{table}

\subsection{$\Gamma$ and $T$ matrices}

The synthetic $\Gamma$ matrix is calculated from
\begin{equation}
\Gamma(x|x') = \sum_{\lambda_1 \cdots \lambda_n} 
(L_1^{-1})_{x_1'  \lambda_1} \cdots (L_n^{-1})_{x_n'  \lambda_n}   
\  {\rm tr} (E_x \, \rho_{\lambda_1} \! \cdots \rho_{\lambda_n}) ,
\label{synthetic gamma}
\end{equation}
using the $L$ matrices from Table~\ref{simulated L table}. The {\it exact} error model is used to calculate ${\rm tr} (E_x \, \rho_{\lambda_1} \! \cdots \rho_{\lambda_n})$. The synthetic $T$ matrix is calculated from 
\begin{equation}
T(x|x') = {\rm tr} (E_x \, \rho_{x'}),
\label{synthetic T}
\end{equation}
where $E_x$ and $\rho_{x'}$ are obtained from the exact error model.

\subsection{Mermin polynomial}

Finally, we simulate the measurement of the Mermin polynomial (\ref{defM4}) in the presence of the noisy multiqubit POVM from step A. We assume that the noisy GHZ state (\ref{defRhoNoisy}) has been prepared in the register and we calculate both the noisy and measurement-error-free values of $M_4$. Then we perform TMEM with the synthetic $\Gamma$ and $T$ matrices, and compare with the exact values. We do this as a function of GHZ state depolarization error $\eta$, with the results of the first of 16 random POVM instances summarized in Figs.~\ref{M4 E2E figure} and \ref{M4 E2E difference figure}.

Figure~\ref{M4 E2E figure} contains four curves. The solid red curve  is the exact value of $M_4$ in the state (\ref{defRhoNoisy}) with no measurement error. The dotted purple curve is the raw data value of $M_4$ that would be measured in the presence of the noisy POVM from the exact error model. It is obtained by including the noisy POVM in each expectation value measurement in (\ref{defM4}). The dashed purple curve is the result of applying TMEM with the synthetic $T$ matrix. It overestimates $M_4$ because the $T$ matrix is corrupted by the presence of imperfect state preparation. The solid purple curve (obscured by the solid red line) is the result of applying TMEM with the synthetic $\Gamma$ matrix. The results in Table~\ref{mermin simulation table} agree with the $\eta=0.2$ case in Fig.~\ref{M4 E2E figure}. The differences between the exact and $\Gamma$-corrected values are too small to be seen in Fig.~\ref{M4 E2E figure}, but are directly plotted in Fig.~\ref{M4 E2E difference figure} on a logarithmic scale.

\begin{figure}
\includegraphics[width=12.0cm]{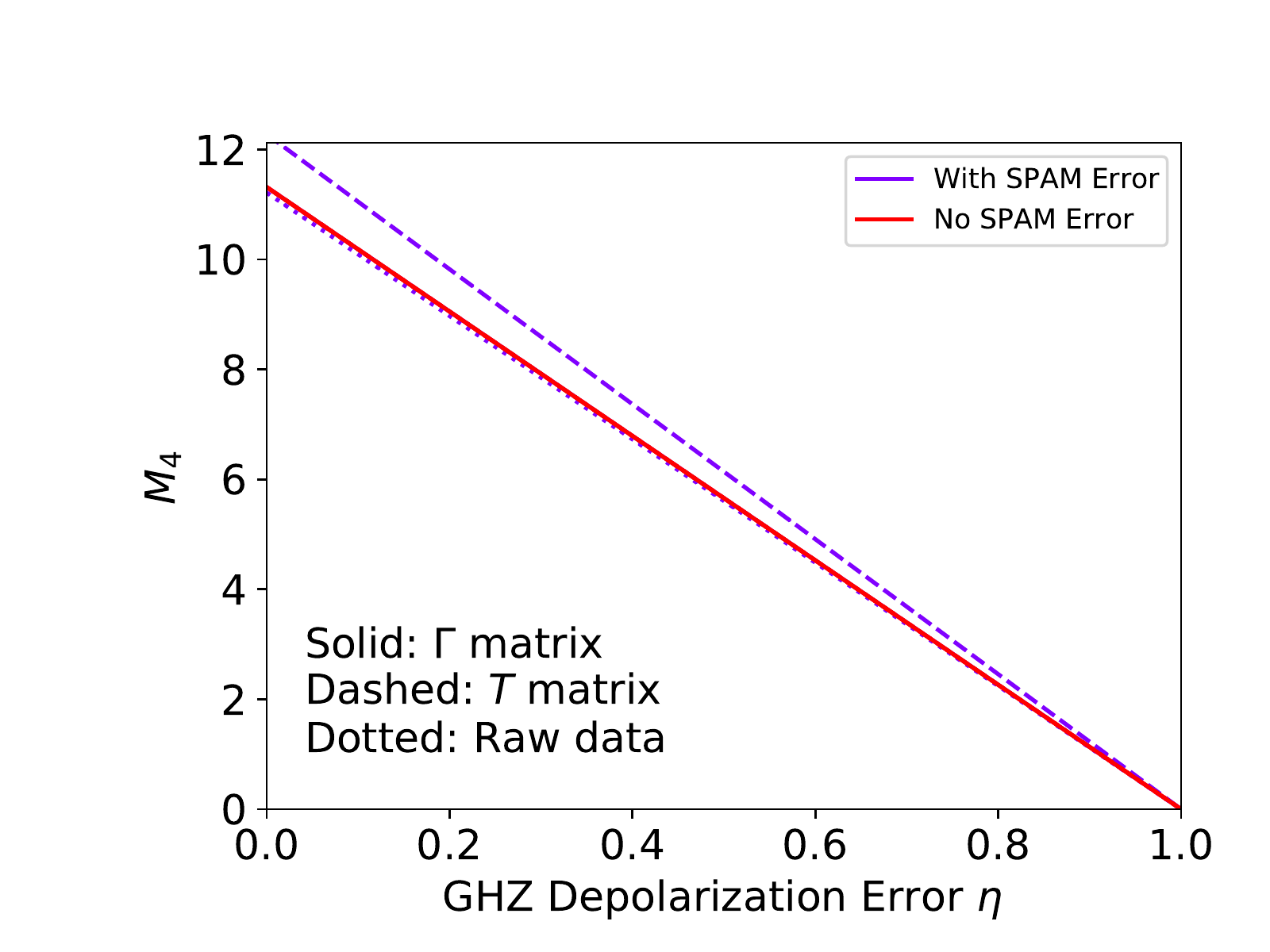} 
\caption{Result of TMEM applied to noisy Mermin polynomial measurement.}
\label{M4 E2E figure}
\end{figure} 

\begin{figure}
\includegraphics[width=12.0cm]{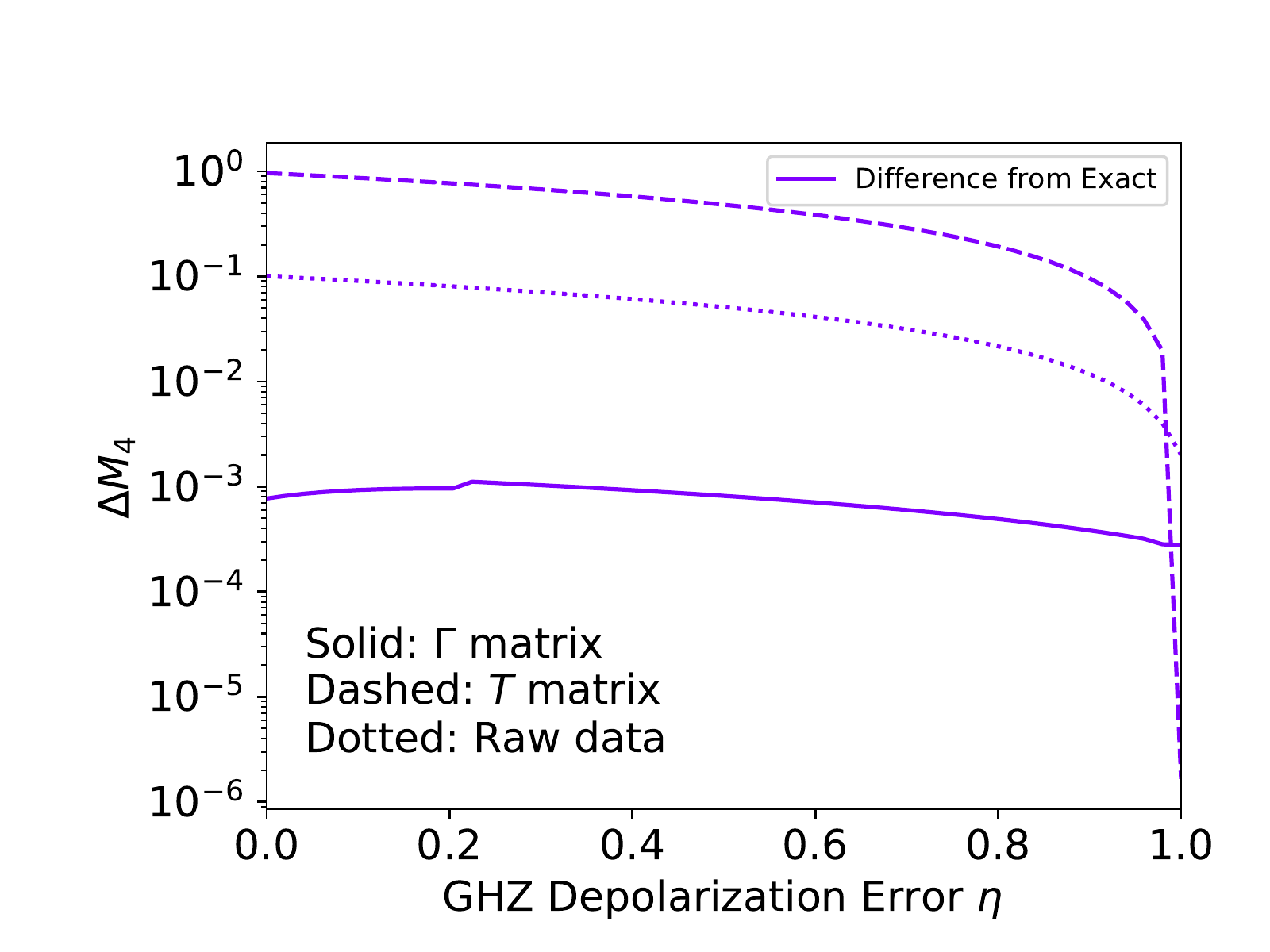} 
\caption{Data from Fig.~\ref{M4 E2E figure} replotted as magnitudes of differences of raw and corrected values from the exact measurement-error-free value.}
\label{M4 E2E difference figure}
\end{figure}

\end{widetext}

\end{document}